\documentclass[preprint,showkeys,amsmath,amssymb]{revtex4}

\usepackage[american]{babel}

\usepackage[dvips]{graphicx}
\usepackage{dcolumn}
\usepackage{bm}   
\usepackage{amssymb, amsmath}
\usepackage{color}

\DeclareSymbolFont{AMSb}{U}{msb}{m}{n}
\DeclareMathSymbol{\R}{\mathbin}{AMSb}{"52}

\newtheorem{theorem}{Theorem}[section]

\usepackage{amssymb, amsmath}
\usepackage{mathrsfs}			

\newcommand{\pf} {{\it Proof.}}

 \newcommand{\ve}[1]{{\bf #1}}
 \newcommand{\vve}[1]{{\bf #1}}
 \newcommand{\veg}[1]{{\boldsymbol {#1}}}	
 \newcommand{\vveg}[1]{{\boldsymbol {#1}}}	

\newcommand{\matthree}[9]{\bracket{\begin{array}{ccc}
		#1	&#2	&#3	\\
		#4	&#5	&#6	\\
		#7	&#8	&#9
		\end{array}}}

\newcommand{\mat}[4]{\bracket{\begin{array}{cc}
		#1	&#2\\
		#3	&#4	   	
		\end{array}}}

\newcommand{\D}{\partial}

\newcommand{\Dt}[1]{\frac {\D #1} {\D t}}

\newcommand{\dt}[1]{\frac {d #1} {d t}}
\newcommand{\delt}{{\Delta t}}
\newcommand{\hot}{o(\delt)}
\newcommand{\delw}{{\Delta\ve w}}

\newcommand{\dms}{n}		
\newcommand{\ntime}{K}		

\newcommand{\DI}[1]{\frac {\D #1} {\D x_1}}	
\newcommand{\DII}[1]{\frac {\D #1} {\D x_2}}

\newcommand{\Dn}[1]{\frac {\D #1} {\D x_\dms}}

\newcommand{\Di}[1]{\frac {\D #1} {\D x_i}}

\newcommand{\DiDj}[1]{\frac {\D^2 #1} {\D x_i \D x_j}}
\newcommand{\DIDI}[1]{\frac {\D^2 #1} {\D x_1^2}}

\newcommand{\excl}[1]{{\backslash \hspace{-0.3em} #1}}
\newcommand{\subA} {{1...r}}			
\newcommand{\subB} {{r+1,...s}}			
\newcommand{\others} {{s+1,...\dms}}		
\newcommand{\subAother} {{1,..r, s+1,..\dms}}	
\newcommand{\subBother} {{r+1,...\dms}}		

\newcommand{\bracket}[1]{\left[#1\right]}

\newcommand{\parenth}[1]{\left(#1\right)}

\DeclareSymbolFont{AMSb}{U}{msb}{m}{n}
\DeclareMathSymbol{\R}{\mathbin}{AMSb}{"52}

\begin{document}
\preprint{Preprint submitted to {Entropy--Complexity}}
\title{The causal interaction between the subnetworks of a complex network}

\author{X. San Liang}
\email{sanliang@courant.nyu.edu}
\affiliation{Fudan University, Shanghai 200438, China}
\affiliation{Shanghai Qizhi Institute (Andrew C. Yao Institute for Artificial
Intelligence), Shanghai 200232, China}

\date{November 25, 2021}

\begin{abstract}
{
Information flow provides a natural measure for the causal interaction between
dynamical events. This study extends our previous rigorous formalism of
componentwise information flow to the {\it bulk} information flow 
between two complex subsystems of a large-dimensional parental system, in
order to investigate problems such as the effective connectivity between
two brain regions, each with millions of neurons involved.
Analytical formulas have been obtained in a closed form. 
Under a Gaussian assumption, their maximum likelihood estimators 
have also been obtained. These formulas have been validated using different
subsystems with preset relations, and they yield causalities just as expected.
On the contrary, the commonly used proxies for the characterization of 
subsystems, such as averages and principal components, 
generally do not work correctly. 
This study can help diagnose the emergence of patterns in complex
systems (e.g., the human brain), and is expected to have applications in many real world problems
in different disciplines such as neuroscience, climate science, fluid dynamics,
financial economics, to name a few.
}
\end{abstract}


\maketitle


\section{Introduction}

When investigating the properties of a complex system, often it is needed to study the interaction between
one subsystem and another subsystem, which themselves also form complex
systems, usually with a large number of components involved. In climate
science, for example, there is much interest in understanding
how one sector of the system 
collaborates with another sector to cause the climate change (see
\cite{IPCC2021} and the references therein); in
neuroscience, it is important to investigate the effective connectivity
from one brain region to another, each with millions of neurons involved
(e.g., \cite{Friston2003}, \cite{Li2011}), 
and the interaction between structures 
(e.g., \cite{Friston2014}, \cite{Friston1995}, \cite{Poo2019}; see more
references in a recent review \cite{Poo2020} ).
This naturally raises a question: How can we study the interaction between 
two subsystems in a large parental system?

An immediate answer coming to mind might be to study the componentwise interactions 
by assessing the causalities between the respective components using, 
say, the classical causal inference approaches (e.g., \cite{Granger},
\cite{Pearl}, \cite{Imbens}).
This is generally infeasible if the dimensionality is large.
For two subsystems, each with, say, 1000 components,
it ends up with 1 million causal relations, making it impossible to analyze, albeit with
all the details. In this case, the details do not make a benefit;
they need to be re-analyzed for a big, interpretable picture of the phenomena. 
On the other hand, in many situations, this is not necessary; 
one needs only a ``bulk'' description of the
subsystems and their interactions. Such examples are seen from the Reynolds
equations for turbulence (e.g., \cite{Batchelor}) and the thermodynamic
description of molecular motions (e.g., \cite{Landau}).
In some fields (e.g., climate science, neuroscience, geography,
etc.), a common practice is simply to take respective averages and
form the mean properties, and to study the interactions between the
proxies, i.e., the mean properties. A more sophisticated way is to extract
the respective principal components (PCs) 
(e.g., \cite{Preisendorfer1998}, \cite{Friston1993_PCA},
\cite{Friston2000_PCA}), 
based on which the interactions
are analyzed henceforth. These approaches, as we will be examining in this
study, however may not work satisfactorily; their validities need to be 
carefully checked before put to application.


During the past 16 years, it has been gradually realized that causality 
in terms of
information flow (IF) is a real physical notion that can be rigorously derived
from first principles (see \cite{Liang2016}). 
When two processes interact, IF provides not
only the direction, but also the strength, of the interaction.
So far the formalism of the IF between two components has 
been well established; see \cite{LK2005}, \cite{Liang2008},
\cite{Liang2014}, \cite{Liang2016}, \cite{Liang2021a}, among others.
It has been shown promising to extend the formalism to subspaces with many
components involved. A pioneering effort is \cite{Majda2007},
where the authors show that the heuristic argument in \cite{LK2005} equally
applies to that between subsystems in the case with only one-way causality.
The recent study on the role of individual nodes in a complex network,
\cite{Liang2021b}, may be viewed as another effort.
(Causality analyses between subspaces with the classical
approaches are rare; a few examples are \cite{Sadoon2019},
\cite{Triacca2018}, etc.)
But a rigorous formalism for more generic problems (e.g., with mutual
causality involved) is yet to be implemented.
This makes the objective of this study, i.e., to study
the interactions between two complex subsystems within 
a large parental system, by investigating
the ``bulk'' information flow between them. 


The rest of the paper is scheduled as follows. 
In section~\ref{sect:derivation}, 
we first present the setting for the problem, 
then derive the IF formulas. Maximum likelihood estimators of
these formulas are given in section~\ref{sect:mle},
which is followed by a validation. 
Finally section~\ref{sect:summary} summarizes the study.

\section{Information flow between two subspaces of a complex system}
	\label{sect:derivation}

Consider an $\dms$-dimensional dynamical system
   \begin{eqnarray}
	&& A:\
	\left\{\begin{array}{l}
	\dt{x_{1}} = F_{1}(x_1,x_2,...,x_\dms; t) 
		+ \sum_{k=1}^m b_{1,k}(x_1,x_2,...,x_\dms; t) \dot w_k\\
	\quad\vdots	\qquad\qquad\qquad\qquad\qquad\quad\ \ \vdots \\
	\dt{x_{r}} = F_{r}(x_1,x_2,...,x_\dms; t) 
		+ \sum_{k=1}^m b_{r,k}(x_1,x_2,...,x_\dms; t) \dot w_k\\
	\end{array}\right. 	\label{eq:gov1}	\\
	&& B:\
	\left\{\begin{array}{l}
	\dt{x_{r+1}} = F_{r+1}(x_1,x_2,...,x_\dms; t) 
		+ \sum_{k=1}^m b_{r+1,k}(x_1,x_2,...,x_\dms; t) \dot w_k\\
	\quad\vdots	\qquad\qquad\qquad\qquad\qquad\quad\ \ \vdots \\
	\dt{x_{s}} = F_{s}(x_1,x_2,...,x_\dms; t) 
		+ \sum_{k=1}^m b_{s,k}(x_1,x_2,...,x_\dms; t) \dot w_k\\
	\end{array}\right. 	\label{eq:gov2} \\
	&&\qquad 
	\left\{\begin{array}{l}
           \dt{x_{s+1}} = F_{s+1}(x_1,x_2,...,x_\dms; t) 
		+ \sum_{k=1}^m b_{s+1,k}(x_1,x_2,...,x_\dms; t) \dot w_k\\
           \quad\vdots	\qquad\qquad\qquad\qquad\qquad\quad\ \ \vdots \\
           \dt{x_\dms} = F_\dms(x_1,x_2,...,x_\dms; t) 
		+ \sum_{k=1}^m b_{\dms k}(x_1,x_2,...,x_\dms; t) \dot w_k.
	\end{array}\right. 	\label{eq:gov3} 
   \end{eqnarray}
where $\ve x \in\R^\dms$ denotes the vector of state variable $(x_1, x_2,..., x_\dms)$, 
$\ve F = (F_1, ..., F_\dms)$ are differentiable functions of $\ve x$ and
time $t$, $\ve w$ is a vector of $m$ independent standard Wiener processes,
and $\vve B = (b_{ij})$ is an $\dms\times m$ matrix of stochastic 
perturbation amplitudes. Here we follow the convention in physics not to
distinguish a random variable from its deterministic counterpart. 
From the components $(x_1,...,x_\dms)$ we separate out two sets,
$(x_1,...,x_r)$ and $(x_{r+1},...,x_s)$, and denote them as
$\ve x_\subA$ and $\ve x_\subB$, respectively. The remaining components 
$(x_{s+1},...,x_\dms)$ are denoted as $\ve x_\others$.
The subsystems formed 
by them are henceforth referred to as $A$ and $B$, and the following
is a derivation of the information flow between them. Note that, for
convenience, here $A$ and $B$ are put adjacent to each other; if not,
the equations can always be rearranged to make them so. 

Associated with (\ref{eq:gov1})-(\ref{eq:gov3}) 
there is a Fokker-Planck equation governing
the evolution of the joint probability density function (pdf) $\rho$ of $\ve x$:
	\begin{eqnarray}	\label{eq:fk}
	\Dt\rho + \DI {\rho F_1} + \DII {\rho F_2} + ... + \Dn{\rho F_n}
	= \frac 12 \sum_{i=1}^\dms \sum_{j=1}^\dms \DiDj {g_{ij}\rho},
	\end{eqnarray}
where $g_{ij} = \sum_{k=1}^m b_{ik} b_{jk}$, $i,j=1,...,\dms$.
Without much loss of generality, $\rho$ is assumed to be compactly
supported on $\R^\dms$.
The joint pdfs of $\ve x_\subA$ and $\ve x_\subB$ are, respectively,
	\begin{eqnarray*}
	&&\rho_\subA = \int_{\R^{\dms-r}} \rho(\ve x) dx_{r+1} ... dx_\dms
	           \equiv \int_{\R^{\dms-r}} \rho(\ve x) d\ve x_\subBother, \\
	&&\rho_\subB = \int_{\R^{\dms-s+r}} \rho(\ve x) dx_1...dx_r
						dx_{s+1}...dx_\dms
	           \equiv \int_{\R^{\dms-s+r}} \rho(\ve x) d\ve x_\subAother.
	\end{eqnarray*}
With respect to them the joint entropies are then
	\begin{eqnarray}
	&&H_A = -\int_{\R^r} \rho_\subA \log\rho_\subA d\ve x_\subA,\\
	&&H_B = -\int_{\R^{s-r}} \rho_\subB \log\rho_\subB d\ve x_\subB.
	\end{eqnarray}
To derive the evolution of $\rho_\subA$, 
integrate out $(x_{r+1},...,x_\dms)$ in (\ref{eq:fk}). 
This yields, by using the assumption of compactness for $\rho$, 
	\begin{eqnarray}	\label{eq:rhoA}
	&& \Dt {\rho_\subA} + 
		\sum_{i=1}^r \Di\ \int_{\R^{\dms-r}} \rho F_i d\ve x_\subBother 
	   = \frac12 \sum_{i=1}^r\sum_{j=1}^r 
	  \int_{\R^{\dms-r}}\DiDj {g_{ij}\rho} d\ve x_\subBother.
	\end{eqnarray}
Similarly, 
	\begin{eqnarray}
	&& \Dt {\rho_\subB} + 
		\sum_{i=r+1}^s \Di\ \int_{\R^{\dms-s+r}} \rho F_i d\ve x_\subAother 
	   = \frac12 \sum_{i=r+1}^s \sum_{j=r+1}^s
	  \int_{\R^{\dms-s+r}}\DiDj {g_{ij}\rho} d\ve x_\subAother.
	\end{eqnarray}
Multiplication of (\ref{eq:rhoA}) by $-(1+\log\rho_\subA)$,
followed by an integration with respect to $\ve x_\subA$ over $\R^r$, we have 
	\begin{eqnarray*}
	&& \dt {H_A} - \sum_{i=1}^r \int_{\R^r} 
	    \bracket{(1+\log\rho_\subA) \cdot \Di\ \int_{\R^{\dms-r}} \rho F_i 
			d\ve x_\subBother}\ d\ve x_\subA  \cr
	&&\qquad =
	  -\frac12 \int_{\R^r}\bracket{ (1+\log\rho_\subA) \cdot  
		\sum_{i=1}^r \sum_{j=1}^r 
		\int_{\R^{\dms-r}} \DiDj {g_{ij}\rho} d\ve x_\subBother
			      }\ d\ve x_\subA.
	\end{eqnarray*}
Note that in the second term of the left hand side, 
the part within the summation is, by integration by parts,
	\begin{eqnarray*}
	&&\int_{\R^r} (\log\rho_\subA) \cdot 
		     \Di\ \parenth{\int_{\R^{\dms-r}} \rho F_i d\ve x_\subBother}
		d\ve x_\subA	\\
	&&\qquad=
	- \int_{\R^r} \int_{\R^{\dms-r}} \rho F_i \Di {\log\rho_\subA}
		d\ve x_\subBother d\ve x_\subA 	\\
	&&\qquad=
	- \int_{\R^\dms} \rho F_i \Di {\log\rho_\subA} \ d\ve x  \\
	&&\qquad=
	-E\bracket{F_i \Di {\log\rho_\subA}}.
	\end{eqnarray*} 
In the derivation, the compactness assumption has been used (variables
vanish at the boundaries).
By the same approach, the right hand side becomes
	\begin{eqnarray*}
	&& -\frac12 \int_{\R^r} \bracket{
		\log\rho_\subA \cdot \sum_{i=1}^r \sum_{j=1}^r
		\int_{\R^{\dms-r}} \DiDj {g_{ij} \rho}
				d\ve x_\subBother	} d\ve x_\subA	\\
	&&\qquad=
	   -\frac12 \sum_{i=1}^r \sum_{j=1}^r \int_{\R^\dms} \bracket{
		\log\rho_\subA \cdot \DiDj {g_{ij} \rho}
					} d\ve x \\
	&&\qquad=
	   -\frac12 \sum_{i=1}^r \sum_{j=1}^r \int_{\R^\dms} \bracket{
		g_{ij} \rho \DiDj {\log\rho_\subA}  
					} d\ve x \\
	&&\qquad=
	   -\frac12 \sum_{i=1}^r \sum_{j=1}^r 
		E \bracket{g_{ij} \DiDj {\log\rho_\subA}  }.
	\end{eqnarray*}
Hence
	\begin{eqnarray}
	\dt {H_A} = - \sum_{i=1}^r E\bracket{F_i \Di {\log\rho_\subA}}
		    - \frac12 \sum_{i=1}^r\sum_{j=1}^r 
			E\bracket{g_{ij} \DiDj {\log\rho_\subA} }.
	\end{eqnarray}
Likewise, we have
	\begin{eqnarray}
	\dt {H_B} = - \sum_{i=r+1}^s E\bracket{F_i \Di {\log\rho_\subB}}
		    - \frac12 \sum_{i=r+1}^s \sum_{j=r+1}^s 
			E\bracket{g_{ij} \DiDj {\log\rho_\subB} }.
	\end{eqnarray}

Now consider the impact of the subsystem $A$ to its peer $B$, written
$T_{A\to B}$. Following Liang (2016)\cite{Liang2016}, 
this is associated with the evolution of the joint entropy of the latter:
	\begin{eqnarray}
	\dt {H_B} = \dt {H_{B\excl A}} + T_{A\to B},
	\end{eqnarray}
where $H_{B\excl A}$ signifies the entropy evolution with the influence of
$A$ excluded, which is found by instantaneously freezing $(x_1,...,x_r)
\equiv \ve x_\subA$ as parameters. To do this, 
examine, on an infinitesimal interval $[t,~t+\delt]$, 
a system modified from the original (\ref{eq:gov1})-(\ref{eq:gov3}) 
by removing the $r$ equations for $x_1$, $x_2$, ..., $x_r$ from the
equation set,
	\begin{eqnarray}
	&&\dt{x_{r+1}} = F_{r+1}(x_1,x_2,...,x_\dms; t) 
		+ \sum_{k=1}^m b_{r+1,k}(x_1,x_2,...,x_\dms; t) \dot w_k\\
	&&\quad\vdots	\qquad\qquad\qquad\qquad\qquad\quad\ \ \vdots \cr
	&&\dt{x_{s}} = F_{s}(x_1,x_2,...,x_\dms; t) 
		+ \sum_{k=1}^m b_{s,k}(x_1,x_2,...,x_\dms; t) \dot w_k\\
	&&\dt{x_{s+1}} = F_{s+1}(x_1,x_2,...,x_\dms; t) 
		+ \sum_{k=1}^m b_{s+1,k}(x_1,x_2,...,x_\dms; t) \dot w_k\\
	&&\quad\vdots	\qquad\qquad\qquad\qquad\qquad\quad\ \ \vdots \cr
	&&\dt{x_\dms} = F_\dms(x_1,x_2,...,x_\dms; t) 
		+ \sum_{k=1}^m b_{\dms k}(x_1,x_2,...,x_\dms; t) \dot w_k.
	\end{eqnarray}
Notice that the $F_i$'s and $b_{ik}$'s still have dependence on
$(x_1,...,x_r)\equiv\ve x_\subA$, which however appear
in the modified system as parameters. By \cite{Liang2016}, 
we can construct a mapping 
$\Phi: \R^{\dms-r} \to \R^{\dms-r}$, 
$\ve x_\excl A(t) \mapsto \ve x_\excl A(t+\delt)$, 
where $\ve x_\excl A$ means $\ve x$ but with $\ve x_\subA$ appearing as
parameters, and study the Frobenius-Perron operator (see, for example, \cite{Lasota1994}) 
of the modified system.
An alternative approach is given by Liang in \cite{Liang2008}, 
which we henceforth follow.
Observe that, on the interval $[t, t+\delt]$, 
corresponding to the modified dynamical system
there is also a Fokker-Planck equation 
	\begin{eqnarray*}	
	&&
	\Dt {\rho_\excl A} + \sum_{i=r+1}^\dms  \Di {F_i\rho_\excl A} 
	= \frac12 \sum_{i=r+1}^\dms \sum_{j=r+1}^\dms 
				\DiDj {g_{ij}\rho_\excl A},\\
	&&
	\rho_\excl A = \rho_\subBother \qquad\qquad\qquad {\rm at\ time\ t.}
	\end{eqnarray*}
Here $g_{ij} = \sum_{k=1}^m b_{ik} b_{jk}$,
$\rho_\excl A$ means the joint pdf of $(x_{r+1},...,x_\dms)$ with 
$\ve x_\subA$ frozen as parameters. 
Note the difference between $\rho_\excl A$ and $\rho_\subBother$; 
the former has $\ve x_\subA$ as parameters, while the latter has no
dependence on $\ve x_\subA$. But they are equal at time $t$.

Integration of the above Fokker-Planck equation with respect to 
$d\ve x_\others$ gives the evolution of the pdf of subsystem $B$ with $A$ frozen
as parameter, written $\rho_{B,\excl A}$:
	\begin{eqnarray}	\label{eq:fk_modified}
	&&
	\Dt {\rho_{B,\excl A}} + \sum_{i=r+1}^s  \int_{\R^{\dms-s}}\Di
{F_i\rho_\excl A} d\ve x_\others
	= \frac12 \sum_{i=r+1}^s \sum_{j=r+1}^s  \int_{\R^{\dms-s}}
				\DiDj {g_{ij}\rho_\excl A} d\ve x_\others,\\
	&&
	\rho_{B,\excl A} = \rho_\subB \qquad\qquad\qquad {\rm at\ time\ t.}
	\end{eqnarray}

Divide (\ref{eq:fk_modified}) by $\rho_{B,\excl A}$, and simplify the
notation $\ve x_\subB$ by $\ve x_B$, to get
	\begin{eqnarray*}
	\Dt {\log\rho_{B,\excl A}} + 
	   \sum_{i=r+1}^s \frac 1{\rho_{B,\excl A}} \int_{\R^{\dms-s}}\Di{F_i\rho_\excl A} d\ve x_\others
	= \frac1{2\rho_{B,\excl A}} 
	  \sum_{i=r+1}^s\sum_{j=r+1}^s \int_{\R^{\dms-s}} \DiDj {g_{ij}\rho_\excl A} d\ve x_\others.
	\end{eqnarray*}
Discretizing, and noticing that $\rho_{B,\excl A}(t) = \rho_\subB(t)$,
we have (in the following, unless otherwise indicated, 
the variables without arguments explicitly specified are assumed to
be at time step $t$)
	\begin{eqnarray*}
	&& \log\rho_{B,\excl A}(\ve x_B; t+\delt)	\\
	&& 
	= \log\rho_\subB(\ve x_B; t) 
	  - \delt \cdot \sum_{i=r+1}^s \frac 1{\rho_\subB} 
		\int_{\R^{\dms-s}} \Di {F_i\rho_\subBother} d\ve x_\others \\
	&&\quad
	  + \frac\delt 2 \sum_{i=r+1}^s \sum_{j=r+1}^s 
			\frac 1 {\rho_\subB}
		\int_{\R^{\dms-s}} \DiDj {g_{ij}\rho_\subBother} d\ve x_\others
	  + \hot.
	\end{eqnarray*}
To arrive at $dH_{B,\excl A}/dt$, we need to find 
	$\log\rho_{B,\excl A}(\ve x_B(t+\delt); t+\delt)$. 
Using the Euler-Bernstein approximation,
	\begin{eqnarray}
	\ve x_B(t+\delt) = \ve x_B(t) + \ve F_B\delt 
		+ \vve B_B \delw
	\end{eqnarray}
where, just like the notation $\ve x_B$, 
	\begin{eqnarray*}
	&& \ve F_B = (F_{r+1},...,F_s)^T,	\\
	&& \vve B_B = \matthree
			   {b_{r+1,1}} \hdots  {b_{r+1,m}}
			   \vdots   \ddots  \vdots
			   {b_{s1}} \hdots  {b_{s m}} \\
	&& \delw = (\Delta w_1, ..., \Delta w_m)^T
	\end{eqnarray*}
and $\Delta w_k \sim N(0, \delt)$, we have 
	\begin{eqnarray*}
	&&\log(\rho_{B,\excl A}(\ve x_B(t+\delt); t+\delt)	\\
	&&
	= \log\rho_\subB(\ve x_B(t) + \ve F_B\delt 
					  + \vve B_B\delw; t)  \\
	&&\ \ \
	  - \delt \cdot \sum_{r+1}^s \frac 1{\rho_\subB} \int_{\R^{\dms-s}} \Di {F_i\rho_\subBother} d\ve x_\others
	  + \frac\delt 2 \sum_{r+1}^s \sum_{r+1}^s \frac 1 {\rho_\subB}
		\int_{\R^{\dms-s}} \DiDj {g_{ij}\rho_\subBother} d\ve x_\others
	  + \hot.	\\
	&&
	= \log\rho_\subB(\ve x_B(t)) 
	  + \sum_{i=r+1}^s \bracket{
	    \Di{\log\rho_\subB} (F_i\delt + \sum_{k=1}^m b_{ik} \Delta w_k)
				   }		\\
	&&\ \ \
	  + \frac12\cdot \sum_{i=r+1}^s \sum_{j=r+1}^s
	    \bracket{
            \DiDj {\log\rho_\subB} (F_i\delt + \sum_{k=1}^m b_{ik}\Delta w_k)
                             \cdot  (F_j\delt + \sum_{l=1}^m b_{jl}\Delta w_l)
		    }		\\
	&&\ \ \
	  - \delt \cdot \sum_{r+1}^s \frac 1{\rho_\subB} 
		\int_{\R^{\dms-s}} \Di {F_i\rho_\subBother} d\ve x_\others
	  + \frac\delt 2 \sum_{r+1}^s \sum_{r+1}^s \frac 1 {\rho_\subB}
		\int_{\R^{\dms-s}} \DiDj {g_{ij}\rho_\subBother} d\ve x_\others
	  + \hot.
	\end{eqnarray*}
Take mathematical expectation on both sides. The left hand side is
$-H_{B, \excl A}(t+\delt)$.
By the Corollary III.I of \cite{Liang2016}, and noting 
$E\Delta w_k = 0$, $E\Delta w_k^2 = \delt$ and the fact that 
$\delw$ are independent of $\ve x_B$, we have
	\begin{eqnarray*}
	&& - H_{B,\excl A}(t+\delt) = -H_B(t) + 
	\delt \cdot E\sum_{i=r+1}^s F_i \Di{\log\rho_\subB}\\
	&&\qquad 
	  + \frac\delt 2 \cdot E \sum_{i=r+1}^s \sum_{j=r+1}^s
	 	\sum_{k=1}^m \sum_{l=1}^m b_{ik} b_{jl} \delta_{kl}
			\DiDj {\log\rho_\subB}	\\
	&&\qquad
	  - \delt \cdot E \sum_{r+1}^s \frac 1{\rho_\subB}
		\int_{\R^{\dms-s}} \Di {F_i\rho_\subBother} d\ve x_\others \\
	&&\qquad
	  + \frac\delt 2 E \sum_{r+1}^s \sum_{r+1}^s \frac 1 {\rho_\subB}
		\int_{\R^{\dms-s}} \DiDj {g_{ij}\rho_\subBother} d\ve x_\others
			 + \hot	\\
	&&=
	-H_B(t) + \delt \cdot E\sum_{i=r+1}^s F_i \Di{\log\rho_\subB}
	  + \frac\delt 2 \cdot E \sum_{i=r+1}^s \sum_{j=r+1}^s
	 	g_{ij} \DiDj {\log\rho_\subB}	\\
	&&\qquad 
	  - \delt \cdot E \sum_{r+1}^s \frac 1{\rho_\subB} 
		\int_{\R^{\dms-s}} \Di {F_i\rho_\subBother} d\ve x_\others \\
	&&\qquad
	  + \frac\delt 2 E \sum_{r+1}^s \sum_{r+1}^s \frac 1 {\rho_\subB}
		\int_{\R^{\dms-s}} \DiDj {g_{ij}\rho_\subBother} d\ve x_\others
			 + \hot.
	\end{eqnarray*}
So
	\begin{eqnarray*}
	&& \dt {H_{B,\excl A}} 
	= \lim_{\delt\to0} \frac {H_{B,\excl A} - H_B(t)} \delt\\
	&&\qquad
	= -E \sum_{i=r+1}^\dms \parenth{F_i \Di{\log\rho_\subB} - 
	     \frac1{\rho_\subB} 
	     \int_{\R^{\dms-s}} \Di{F_i\rho_\subBother} d\ve x_\others }  \\
	&&\qquad\quad
	  -\frac12 E \sum_{i=r+1}^s \sum_{j=r+1}^s
		\parenth{g_{ij} \DiDj {\log\rho_\subB} + 
		\frac1 {\rho_\subB} \DiDj \
		\int_{\R^{\dms-s}} {g_{ij}\rho_\subBother} d\ve x_\others }.
	\end{eqnarray*}
Hence the information flow from $\ve x_\subA$ to $\ve x_\subB$ is
	\begin{eqnarray*}
	&&T_{A\to B} = \dt {H_B} - \dt {H_{B,\excl A}}	\cr
	&&\qquad
	=-E \sum_{i=r+1}^s \parenth{F_i \Di {\log\rho_\subB}  }
	 -\frac12 E \sum_{i=r+1}^s \sum_{j=r+1}^s 
		\parenth{g_{ij} \DiDj {\log\rho_\subB}  }	\cr
	&&\qquad\ \ \ \
	 +E \sum_{i=r+1}^s \parenth{F_i \Di {\log\rho_\subB} 
		- \frac 1 {\rho_\subB} \int_{\R^{\dms-s}} 
			\Di {F_i \rho_\subBother} d\ve x_\others}  \cr
	&&\qquad\ \ \ \
	 + \frac12 E \sum_{i=r+1}^s \sum_{j=r+1}^s 
	   \parenth{g_{ij}\DiDj {\log\rho_\subB} 
	  + \frac1{\rho_\subB} \int_{\R^{\dms-s}} 
		\DiDj {g_{ij}\rho_\subBother} d\ve x_\others } \cr
	&&\qquad
	= -E\bracket{\sum_{i=r+1}^s \frac1 {\rho_\subB}
		  \int_{\R^{\dms-s}}  \Di{F_i\rho_\subBother} d\ve x_\others}\cr
	&&\qquad\ \ \ \
	  + \frac12 E\bracket{
		\sum_{i=r+1}^s \sum_{j=r+1}^s \frac 1 {\rho_\subB}
	    \int_{\R^{\dms-s}} \DiDj {g_{ij} \rho_\subBother} d\ve x_\others
			     }.
	\end{eqnarray*}

Likewise, we can get the information flow from subsystem $B$ to subsystem $A$.
These are summarized in the following theorem.
	\begin{theorem}
	For the dynamical system (\ref{eq:gov1})-(\ref{eq:gov3}), if the probability 
	density function (pdf) of $\ve x$ is compactly supported, then
	the information flow from $\ve x_\subA$ to $\ve x_\subB$, and
	that from $\ve x_\subB$ to $\ve x_\subA$ are (in nats per unit
	time), respectively,
	\begin{eqnarray}	\label{eq:TAB}
	&& T_{A\to B} 
	= -E\bracket{\sum_{i=r+1}^s \frac1 {\rho_\subB}
		\int_{\R^{\dms-s}} \Di{F_i\rho_\subBother} d\ve x_\others } \cr
	&&\qquad\ \ \ \
	  + \frac12 E\bracket{
		\sum_{i=r+1}^s \sum_{j=r+1}^s \frac 1 {\rho_\subB}
	        \int_{\R^{\dms-s}} \DiDj {g_{ij} \rho_\subBother} d\ve x_\others
			     },
	\end{eqnarray}
	\begin{eqnarray}	\label{eq:TBA}
	&&T_{B\to A} 
	= -E\bracket{\sum_{i=1}^r \frac1 {\rho_\subA}
		\int_{\R^{\dms-s}} \Di{F_i\rho_\subAother} d\ve x_\others } \cr
	&&\qquad\ \ \ \
	  + \frac12 E\bracket{
		\sum_{i=1}^r \sum_{j=1}^r \frac 1 {\rho_\subA}
		\int_{\R^{\dms-s}} \DiDj {g_{ij} \rho_\subAother} d\ve x_\others
			     },
	\end{eqnarray}
	where $g_{ij} = \sum_{k=1}^m b_{ik} b_{jk}$,
	and $E$ denotes mathematical expectation.
	\end{theorem}

When $r=1$, $s=\dms=2$, (\ref{eq:TBA}) reduces to 
	\begin{eqnarray*}
	T_{B\to A} = -E\bracket{\frac1 {\rho_1} \DI {F_1\rho_1} }
		   + \frac12 E\bracket{\frac1 {\rho_1} \DIDI {g_{11}\rho_1}  }
	\end{eqnarray*}
which is precisely the same as the Eq.~(15) in \cite{Liang2008}; same
holds for (\ref{eq:TAB}). These equations are hence verified.

The following theorem forms the basis for causal inference.
	\begin{theorem}		\label{thm:PNC}
	If the evolution of subsystem $A$ (resp. $B$) does not depend
	on $\ve x_\subB$ (resp. $\ve x_\subA$), then
	$T_{B\to A}=0$ (resp. $T_{A\to B}=0$).
	\end{theorem}
\pf 
We only check the formula for $T_{B\to A}$.
	In (\ref{eq:TBA}), the deterministic part
	\begin{eqnarray*}
	&&
	-E\bracket{\sum_{i=1}^r \frac1 {\rho_\subA}
	\int_{\R^{\dms-s}} \Di{F_i\rho_\subAother} d\ve x_\others } \cr
	&& 
	= -\sum_{i=1}^r \int_{\R^r} \int_{\R^{s-r}}
	   \bracket{\parenth{\frac {\rho_{1,...,s}} {\rho_\subA} }
		    \int_{\R^{\dms-s}} \Di {F_i\rho_\subAother} d\ve x_\others
		   } d\ve x_\subA d\ve x_\subB.
	\end{eqnarray*}
Now $F_i$ is independent of $\ve x_\subB$, and note that $\rho_\subAother$
is also so. Thus we may integrate $\rho_{1,..,s}$ within the parenthesis
directly with respect to $d\ve x_\subB$, yielding
	\begin{eqnarray*}
           {\frac {\int_{\R^{s-r}} \rho_{1,...,s} d\ve x_\subB} 
		          {\rho_\subA} }
	   = \frac {\rho_\subA} {\rho_\subA}
	   = 1.
	\end{eqnarray*}
By the compactness of $\rho$, the whole deterministic part hence vanishes.
Likewise, it can be proved that the stochastic part also vanishes.

This theorem allows us to identify the causality with information flow.
Ideally, if $T_{B\to A}=0$, then $B$ is not causal to $A$, and vice versa;
same holds for $T_{A\to B}$.

\section{Information flow between linear subsystems and its estimation}	\label{sect:mle}

Linear systems provide the simplest framework which is usually taken as the
first step toward a more generic setting. Simple as it may be, it has been
demonstrated in practice that linear results often provide a good
approximation of an otherwise much more complicated problem.
It is thence of interest to examine this special case.

Let 
	\begin{eqnarray}
	F_i = f_i + \sum_{j=1}^\dms a_{ij} x_j,
	\end{eqnarray}
where $f_i$ and $a_{ij}$ are constants. 
Also suppose that $b_{ij}$ are constants; that is to say, the noises are
additive. Then $g_{ij}$ are also constants. Thus, in (\ref{eq:TBA}),
	\begin{eqnarray*}
	&&E\parenth{\frac 1 {\rho_\subA} 
	   \int_{\dms-s} \DiDj {g_{ij} \rho_\subAother } d\ve x_\others  }\\
 	&&\qquad =
           g_{ij} \int_{\R^s} \rho(\ve x_{1...s}) {\frac 1 {\rho_\subA}  
			\DiDj {\int_{} \rho_\subAother } d\ve x_\others } 
				d\ve x_{1...s} \\
	&&\qquad =
	   g_{ij} \int_{\R^r} \int_{\R^{s-r}} \frac {\rho_{1...s}} {\rho_\subA}
			\DiDj {\rho_\subA} d\ve x_\subB d\ve x_\subA  \\
	&&\qquad =
	   g_{ij} \int_{\R^r} 1 \cdot
			\DiDj {\rho_\subA} d\ve x_\subA \\
	&&\qquad =
	  0.
	\end{eqnarray*}
Same holds in (\ref{eq:TAB}).
So the stochastic parts in both (\ref{eq:TAB})  and (\ref{eq:TBA}) vanish.

Since a linear system initialized with a Gaussian process will always be 
Gaussian, we may write
the joint pdf of $\ve x$ as
	\begin{eqnarray}
	\rho(x_1,...,x_\dms) = \frac 1 {\sqrt{(2\pi)^\dms \det\vveg\Sigma}}
	e^{-\frac12 (\ve x - \veg\mu)^T \vveg\Sigma^{-1} (\ve x - \veg\mu)},
	\end{eqnarray}
where $\vveg\Sigma = (\sigma_{ij})_{\dms\times\dms}$ 
is the population covariance matrix of $\ve x$.
By the property of Gaussian process, 
it is easy to show
	\begin{eqnarray}
	\rho_\subB(x_{r+1},...,x_s) 
	= \frac 1 {\sqrt{(2\pi)^{s-r} \det\vveg\Sigma_B}}
	e^{-\frac12 (\ve x_B - \veg\mu_B)^T 
		    \vveg\Sigma_B^{-1} (\ve x_B - \veg\mu_B)},
	\end{eqnarray}
where 
$\ve x_B =(x_{r+1},...,x_s)$,
	$\veg\mu_B = (\mu_{r+1},...,\mu_s)$ is the vector of the means of 
$\ve x_B$, and $\vveg\Sigma_B$ the covariance matrix of $\ve x_B$.
For easy correspondence, we will augment $\ve x_B$, $\veg\mu_B$, 
and $\vveg\Sigma_B$,
so that their entries have the same indices as their counterparts 
in $\ve x$, $\veg\mu$ and $\vveg\Sigma$. 
Separate $F_i$ into two parts
	\begin{eqnarray*}
	F_i 
	&=& \bracket{f_i + \sum_{j=1}^r a_{ij} x_j  
			   + \sum_{j=s+1}^\dms a_{ij} x_j }
		+ \bracket{\sum_{j=r+1}^s a_{ij} x_j }	\\
	&\equiv& F'_i + F''_i,
	\end{eqnarray*}
where $F'_i$ and $F''_i$ correspond to the respective parts in the two
square brackets. So $F'_i$ has nothing to do with the subspace $B$. By 
Theorem~\ref{thm:PNC}, this part does not contribute to the causality 
from $A$ to $B$, so we only need to consider $F''_i$ in evaluating $T_{A\to
B}$; that is to say,
	\begin{eqnarray*}
	T_{A\to B} 
	&=& -E\bracket{\sum_{i=r+1}^s \frac1 {\rho_\subB}
     	   \Di\ {\int_{\R^{\dms-s}} F_i\rho_\subBother d\ve x_\others}}  \\
	&=& -E\bracket{\sum_{i=r+1}^s \frac1 {\rho_\subB}
     	   \Di\ {\int_{\R^{\dms-s}} F''_i\rho_\subBother d\ve x_\others}}\\
	&=& -E\bracket{\sum_{i=r+1}^s \frac1 {\rho_\subB}
     	   \Di{F''_i\rho_\subB }}\\
	&=&
	-\sum_{i=r+1}^s \bracket{E\parenth{F''_i \Di {\log\rho_\subB} } 
				  + E\parenth{\Di {F''_i} }   }.  
	\end{eqnarray*}
The second term in the bracket is $a_{ii}$. The first term is
	\begin{eqnarray*}
	&&F''_i \Di {\log\rho_\subB} = \bracket{\sum_{j=r+1}^s a_{ij} x_j}
	\cdot \Di\ \bracket{-\frac12 (\ve x_B - \veg\mu_B)^T
	\vveg\Sigma_B^{-1} (\ve x_B - \veg\mu_B) }\\
	&&
	= \parenth{\sum_{j=r+1}^s a_{ij} x_j} \dot
	  \sum_{j=r+1}^s \parenth{-\frac {\sigma_{ij}'+\sigma_{ji}'} 2 }
			  \cdot (x_j - \mu_j).
	\end{eqnarray*}
Here $\sigma_{ij}'$ is the $(i,j)^{\rm th}$ entry of the matrix
	\begin{eqnarray*}
	\matthree {\vve I} {\vve 0}   		  {\vve 0}
	  	  {\vve 0} {\vveg\Sigma_{B}^{-1}} {\vve 0}
		  {\vve 0} {\vve 0}		  {\vve I}.
	\end{eqnarray*}
Since here only $1\le i,j \le s$ are in question, this is equal to the 
$(i,j)^{\rm th}$ entry of the matrix
	\begin{eqnarray*}
	\mat {\vve I_{r\times r}} {\vve 0_{r\times (s-r)}}  
	  {\vve 0_{(s-r)\times r}}  {\vveg\Sigma_{B}^{-1}}
	\end{eqnarray*}
As $\vveg\Sigma_B$ is symmetric, so is $\vveg\Sigma_B^{-1}$,
and hence $(\sigma_{ij}'+\sigma_{ji}')/2 = \sigma_{ij}'$.
So
	\begin{eqnarray*}
	&&-E F_i \Di {\log\rho_\subB} = - E \sum_{j=1}^s a_{ij} x_j 
	      \cdot \sum_{j=r+1}^s (-\sigma_{ij}')\cdot (x_j - \mu_j) \\
	&&= E \sum_{k=1}^s a_{ik} (x_k - \mu_k) \cdot
	      \sum_{j=r+1}^s \sigma_{ij}' (x_j - \mu_j)	\\
	&&= \sum_{k=1}^s \sum_{j=r+1}^s a_{ik} \sigma_{ij}' 
			E(x_k - \mu_k) (x_j - \mu_j)	\\
	&&= \sum_{k=1}^s \sum_{j=r+1}^s a_{ik} \sigma_{ij}'\sigma_kj.
	\end{eqnarray*}
Substituting back, we obtain a very simplified result for
$T_{A\to B}$. Likewise $T_{B\to A}$ can also be obtained, as shown in the
following.
	\begin{theorem}
	In (\ref{eq:gov1})-(\ref{eq:gov3}), suppose $b_{ij}$ are
	constants, and
	\begin{eqnarray}
	&& F_i = f_i + \sum_{j=1}^\dms a_{ij} x_j,
	\end{eqnarray}
	where $f_i$ and $a_{ij}$ are also constants. 
	Further suppose that initially $\ve x$ has a Gaussian distribution, 
	then
	\begin{eqnarray}
	T_{A\to B} &=& \sum_{i=r+1}^s
	\bracket {
	\sum_{j=r+1}^s \sigma_{ij}' 
		\parenth{\sum_{k=1}^s a_{ik} \sigma_{kj}}
		- a_{ii}
		 },		\label{eq:TAB_linear}
	\end{eqnarray}
	where $\sigma_{ij}'$ is the $(i,j)^{\rm th}$ entry of
	$\mat {\vve I_{r\times r}} {\vve 0}  
	  {\vve 0}  {\vveg\Sigma_{B}^{-1}}$, and
	\begin{eqnarray}
	T_{B\to A} &=& \sum_{i=1}^r
	\bracket {
	\sum_{j=1}^r \sigma_{ij}''
		\parenth{\sum_{k=1}^s a_{ik} \sigma_{kj}}
		- a_{ii}
		 },		\label{eq:TBA_linear}
	\end{eqnarray}
	where $\sigma_{ij}''$ is the $(i,j)^{\rm th}$ entry of
	$\mat 
	  {\vveg\Sigma_{A}^{-1}}  {\vve 0}
	  {\vve 0}    {\vve I_{(s-r)\times (s-r)}} $.
	\end{theorem}

Given a system like (\ref{eq:gov1})-(\ref{eq:gov3}), we can
evaluate in a precise sense the information flows among the components.
Now suppose, instead of the dynamical system, what we have are just $\dms$ 
time series with $\ntime$ steps, $\ntime\gg\dms$, 
$\{x_1(k)\}, \{x_2(k)\},..., \{x_\dms(k)\}$.
We can estimate the system from the
series, and then apply the information flow formula to fulfill the task.
Assume a linear model as shown above, and assume $m=1$.
following Liang (2014)\cite{Liang2014}, 
the maximum likelihood estimator (mle) of $a_{ij}$
is equal to the least-square solution of the following
over-determined problem
	\begin{eqnarray*}
	\left(\begin{array}{ccccc}
	1  &x_1(1)  &x_2(1) &...  &x_n(1) \\
	1  &x_1(2)  &x_2(2) &...  &x_n(2) \\
	1  &x_1(3)  &x_2(3) &...  &x_n(3) \\
	\vdots &\vdots &\vdots \ddots &\vdots\\
	1  &x_1(\ntime)  &x_2(\ntime) &...  &x_n(\ntime) 
	\end{array}\right) 
		\left(\begin{array}{c}
		f_i \\
		a_{i1}\\
		a_{i2}\\
		\vdots\\
		a_{in}
		\end{array}\right)
	=
		\left(\begin{array}{c}
		\dot x_i(1) \\
		\dot x_i(2) \\
		\dot x_i(3) \\
		\vdots\\
		\dot x_i(\ntime) 
		\end{array}\right)
	\end{eqnarray*}
where $\dot x_i(k) = (x_i(k+1)-x_i(k))/\delt$ 
($\delt$ is the time stepsize), 
for $i=1,2,...,\dms$, $k=1,...,\ntime$.  
Use overbar to denote the time mean over the $\ntime$ steps.
The above equation is
	\begin{eqnarray*}
	\left(\begin{array}{ccccc}
	1  &\bar x_1         &\bar x_2        &...  &\bar x_n    \\
	0  &x_1(2)-\bar x_1  &x_2(2)-\bar x_2 &...  &x_n(2)-\bar x_n \\
	0  &x_1(3)-\bar x_1  &x_2(3)-\bar x_2 &...  &x_n(3)-\bar x_n \\
	\vdots &\vdots &\vdots \ddots &\vdots\\
	0  &x_1(\ntime)-\bar x_1  &x_2(\ntime)-\bar x_2 &...&x_n(\ntime)-\bar x_n
	\end{array}\right) 
		\left(\begin{array}{c}
		f_i \\
		a_{i1}\\
		a_{i2}\\
		\vdots\\
		a_{i\dms}
		\end{array}\right)
	=
		\left(\begin{array}{c}
		\bar{\dot x}_i \\
		\dot x_i(2) - \bar{\dot x}_i \\
		\dot x_i(3) - \bar{\dot x}_i \\
		\vdots\\
		\dot x_i(\ntime) - \bar{\dot x}_i
		\end{array}\right)
	\end{eqnarray*}
Denote by $\vve R$ the matrix
	$$\left(\begin{array}{ccccc}
	x_1(2)-\bar x_1  &x_2(2)-\bar x_2 &...  &x_n(2)-\bar x_n \\
	\vdots &\vdots &\vdots \ddots &\vdots\\
	x_1(\ntime)-\bar x_1  &x_2(\ntime)-\bar x_2 &...&x_n(\ntime)-\bar x_n
	\end{array}\right) ,$$
$\ve q$ the vector 
$(x_i(2)-\bar{\dot x}_i, ..., x_i(\ntime)-\bar{\dot x}_i)^T$,
and $\ve a_i$ the row vector $(a_{i1},...,a_{i\dms})^T$.
Then $\vve R \ve a_i = \ve q$. The least square solution of $\ve a_i$,
$\ve {\hat a}_i$, solves 
	\begin{eqnarray*}
	\vve R^T \vve R \ve {\hat a}_i = \vve R^T \ve q.
	\end{eqnarray*}
Note $\vve R^T \vve R$ is $\ntime\vve C$, where $\vve C=(c_{ij})$ 
is the sample covariance matrix. So
	\begin{eqnarray}
		\left(\begin{array}{c}
		\hat a_{i1}\\
		\hat a_{i2}\\
		\vdots\\
		\hat a_{i\dms}
		\end{array}\right)
	= \vve C^{-1} 
		\left(\begin{array}{c}
		 c_{1,di}\\
		 c_{2,di}\\
		\vdots\\
		 c_{\dms,di}
		\end{array}\right)
	\end{eqnarray}
where $c_{j,di}$ is the sample covariance between the series $\{x_j(k)\}$  and 
$\{ (x_i(k+1) - x_i(k))/\delt \}$.

So finally, the mle of $T_{A\to B}$ is 
	\begin{eqnarray}	\label{eq:TAB_est}
	\hat T_{A\to B} 
	= \sum_{i=r+1}^s \bracket{
	  \sum_{j=r+1}^s c_{ij}' \parenth{\sum_{k=1}^s \hat a_{ik} c_{kj}}
		- \hat a_{ii}
				  },
	\end{eqnarray}
where $c_{ij}'$ is the $(i,j)^{\rm th}$ entry of $\tilde{\vve C}^{-1}$, and
	\begin{eqnarray}
	\tilde {\vve C} =
	\left(\begin{array}{cc}
	 {\vve I_{r\times r}}       & {\vve 0}_{r\times(s-r) }  \\
	  {\vve 0_{(s-r) \times r}} & 
			{\matthree {c_{r+1,r+1}} {...}  {c_{r+1,s}}
			\vdots			\vdots   \vdots
		        {c_{s,r+1}}  {...}  {c_{s,s}} 
			}
	\end{array}\right). 
	\end{eqnarray}

Likewise, 
	\begin{eqnarray}	\label{eq:TBA_est}
	\hat T_{B\to A} 
	&=& \sum_{i=1}^r \bracket{
	  \sum_{j=1}^r c_{ij}'' \parenth{\sum_{k=1}^s \hat a_{ik} c_{kj}}
		- \hat a_{ii}
				  }	\\
	\end{eqnarray}
Here
	\begin{eqnarray}
	\tilde{\tilde {\vve C}} =
	\left(\begin{array}{cc}
	{\matthree {c_{11}} {...}  {c_{1r}}	
	\vdots			\vdots   \vdots
	{c_{r1}}  {...}  {c_{rr}} 
	}				& {\vve 0_{r\times(s-r)}} 	\\
	{\vve 0_{(s-r)\times r}}	& {\vve I_{(s-r)\times(s-r)}}
	\end{array}\right),
	\end{eqnarray}
and $c''_{ij}$ is the $(i,j)th$ entry of ${\tilde{\tilde {\vve C}}}^{_-1}$.

When $n=2$ and $r=1$, hence $s=1$,
$\tilde{\tilde {\vve C}} = \mat {c_{11}} 0 0 1$, so
$c''_{11} = c_{11}^{-1}$.  Eq.~(\ref{eq:TBA_est}) hence becomes 
	\begin{eqnarray*}
	\hat T_{B\to A}
	&=& c_{11}'' \sum_{k=1}^2 \hat a_{1k} c_{k1} - \hat a_{11} \\
	&=& \frac 1 {c_{11}} (\hat a_{11} c_{11} 
	  + \hat a_{12} c_{12}) - \hat a_{11}  \\
	&=&
	\frac {c_{11} c_{12} c_{2,d1} - c_{12}^2 c_{1,d1}}
		{c_{11}^2 - c_{11} c_{12}^2},
	\end{eqnarray*}
recovering the well-known Eq.~(10) in \cite{Liang2014}.

\section{Validation}
\subsection{One-way causal relation}

\begin{figure}[h]
\begin{center}
  \includegraphics[width=0.5\textwidth]{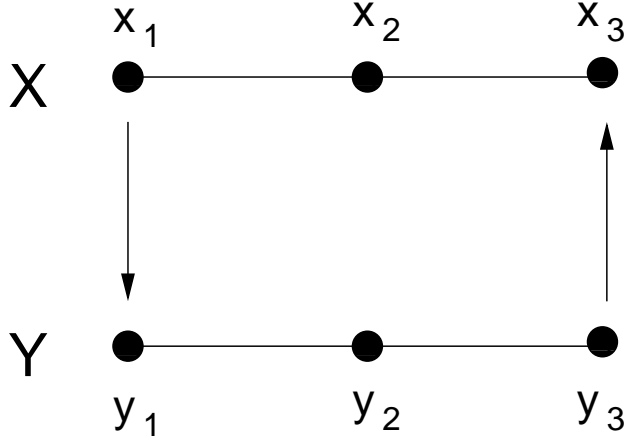}
  \caption{The preset coupling between the toy ``brain regions'' 
	   or subsystems $X$ and $Y$.
	}\label{fig:coupling}
  \end{center}
\end{figure}

To see if the above formalism works, consider the vector autoregressive
(VAR) process, mimicking a noisy toy ``brain'' with six neurons:
	\begin{eqnarray}
	&&X:\ 
	\left\{\begin{array}{l}
	x_1(n+1) = -0.5 x_1(n) + 0.5 x_2(n) + 0.2 x_3(n) + e_{x1}(n+1),\\
	x_2(n+1) = 0 x_1(n) - 0.2 x_2(n) - 0.6 x_3(n) + e_{x2}(n+1), \\
	x_3(n+1) = -0.2 x_1(n) + 0.4 x_2(n) - 0.2 x_3(n) 
		   + \varepsilon_3 y_3(n)
		   + e_{x3}(n+1),
	\end{array}\right.		\\
	&&Y:\ 
	\left\{\begin{array}{l}
	y_1(n+1) = -0.2 y_1(n) - 0.5 y_2(n) + 0 y_3(n)
		   - \varepsilon_1 x_1(n)
		   + e_{y1}(n+1),			\\	
	y_2(n+1) = 0.5 y_1(n) - 0.6 y_2(n) + 0.4 y_3(n) + e_{y2}(n+1),\\
	y_3(n+1) = -0.1 y_1(n) - 0.4 y_2(n) - 0.5 y_3(n) + e_{y3}(n+1)
	\end{array}\right.,
	\end{eqnarray} 
where $e_{xi}, e_{yi} \sim N(0,1)$, $i=1,2,3$, are independent. 
As schematized in Fig.~\ref{fig:coupling}, 
$(x_1,x_2,x_3)$ and $(y_1,y_2,y_3)$ form two subsystems (two ``brain
regions''), written $X$ and $Y$, respectively.
They are coupled only through the first and third components; more
specifically, $x_1$ drives $y_1$, and $Y$ feedbacks to $X$ through
coupling $y_3$ with $x_3$. The strength of the coupling is determined by
the parameters $\varepsilon_1$ and $\varepsilon_3$.
In this subsection, $\varepsilon_3=0$, so the causality is one-way, i.e.,
from $X$ to $Y$ without feedback.

\begin{figure}[h]
\begin{center}
  \includegraphics[width=0.45\textwidth]{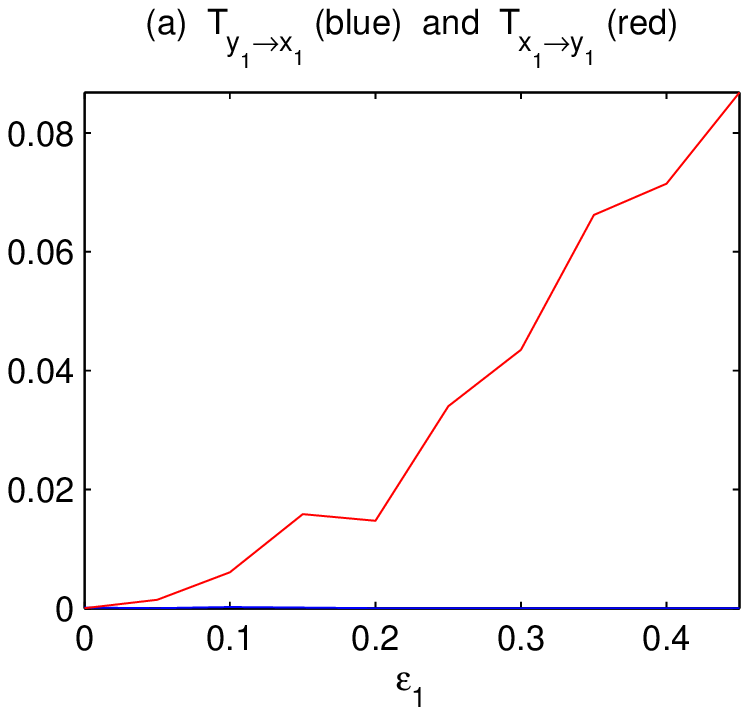}
  \includegraphics[width=0.45\textwidth]{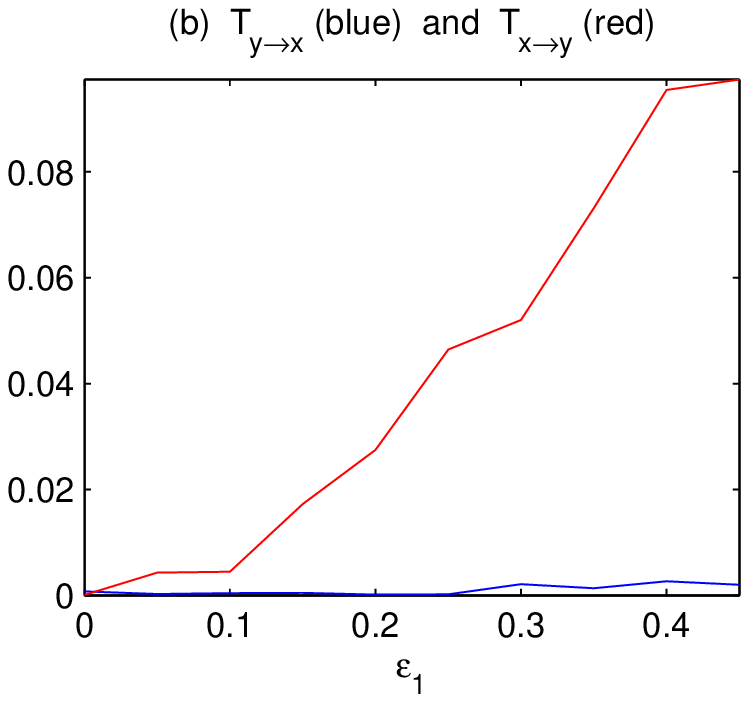}
  \includegraphics[width=0.45\textwidth]{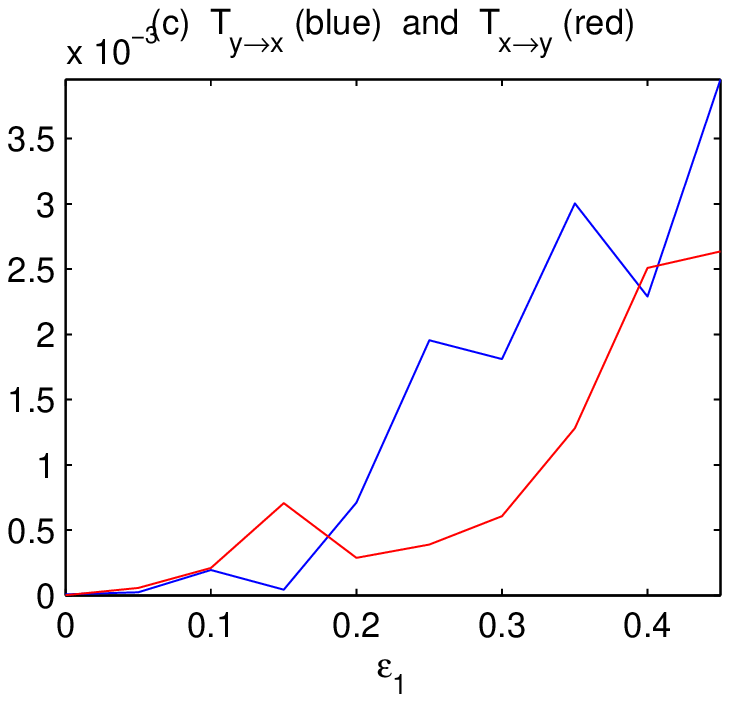}
  \includegraphics[width=0.45\textwidth]{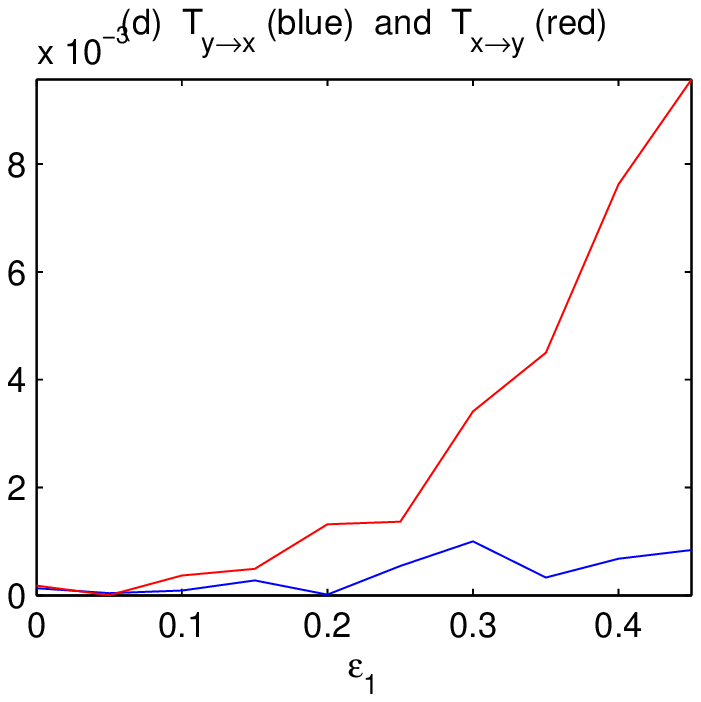}
  \caption{The absolute information flows between subspaces $X$ and $Y$ 
	as functions of the coupling coefficients $\varepsilon_1$
	($\varepsilon_3=0$). (a) The componentwise information flows
	  between $x_1$ and $y_1$; (b) the bulk information flows
	  between subsystems $X$ and $Y$ computed with (\ref{eq:TAB_est})
	  and (\ref{eq:TBA_est}); (c) the information flows between 
	  $\bar x$ and $\bar y$; (d) the information flows between the 
	  first principal components of ($x_1,x_2,x_3$) 
	  and ($y_1,y_2,y_3$), respectively (units: nats per time step).
	}\label{fig:T1}
  \end{center}
\end{figure}

Initialized with random numbers, we iterate the process for 20,000 steps, 
and discard the first 10,000 steps to form six time series with 
a length of 10,000 steps. Using the algorithm by 
Liang (e.g.,\cite{Liang2008}, \cite{Liang2014}, \cite{Liang2016}, \cite{Liang2021a}), 
the information flows between $x_1$ and $y_1$ can be rather accurately
obtained. As shown in Fig.~\ref{fig:T1}a, the information flow/causality
from $X$ to $Y$ increases with $\varepsilon_1$, and there is no causality
the other way around, just as expected.
Since there is no other coupling existing, 
one can imagine that the bulk information flows must also bear a
similar trend. Using (\ref{eq:TAB_est}) and (\ref{eq:TBA_est}), 
the estimators are indeed like that, as shown in Fig.~\ref{fig:T1}b.
This demonstrates the success of the above formalism.

Since practically averages and principal components (PCs) have been widely 
used to measure the complex subsystem variations, we also compute the
information flows between $\bar x = \frac13 (x_1+x_2+x_3)$ 
and $\bar y = \frac13 (y_1+y_2+y_3)$, and that between the
first PCs of $(x_1,x_2,x_3)$ and $(y_1,y_2,y_3)$. The
results are plotted in Figs.~\ref{fig:T1}c and d, respectively.
As can be seen, the principal component analysis (PCA) method works 
just fine in this case. By comparison the averaging method yields
an incorrect result.

The incorrect inference based on averaging is within expectation. In a
network with complex causal relations, for example, with a causality from
$y_2$ to $y_1$, the averaging of $y_1$ with $y_2$ is equivalent to mixing
$y_1$ with its future state, which is related to the contemporary state of 
$x_1$, and hence will yield a spurious causality to $x_1$. The PCA
here functions satisfactorily perhaps because it, 
in selecting the most coherent structure, discards most of the
influences from other (implicit) time steps. But the relative success
of PCA may not be robust, as evidenced in the following mutually causal case.

\subsection{Mutually causal relation}

\begin{figure}[h]
\begin{center}
  \includegraphics[width=0.9\textwidth]{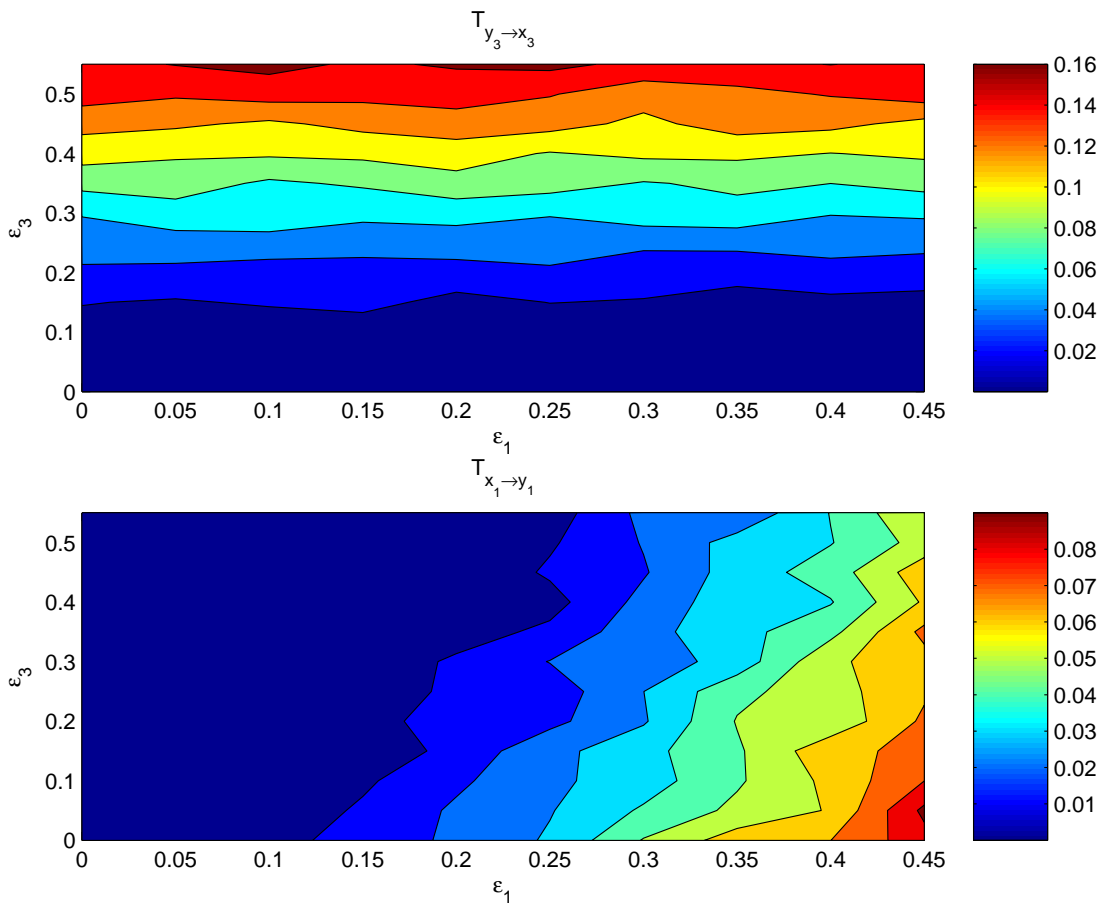}
  \caption{The absolute information flow from $y_3$ to $x_3$ and 
	that from $x_1$ to $y_1$
	as functions of $\varepsilon_1$ and $\varepsilon_3$. 
	The units are in nats per time step.
	}\label{fig:T_individual}
  \end{center}
\end{figure}

\begin{figure}[h]
\begin{center}
  \includegraphics[width=0.9\textwidth]{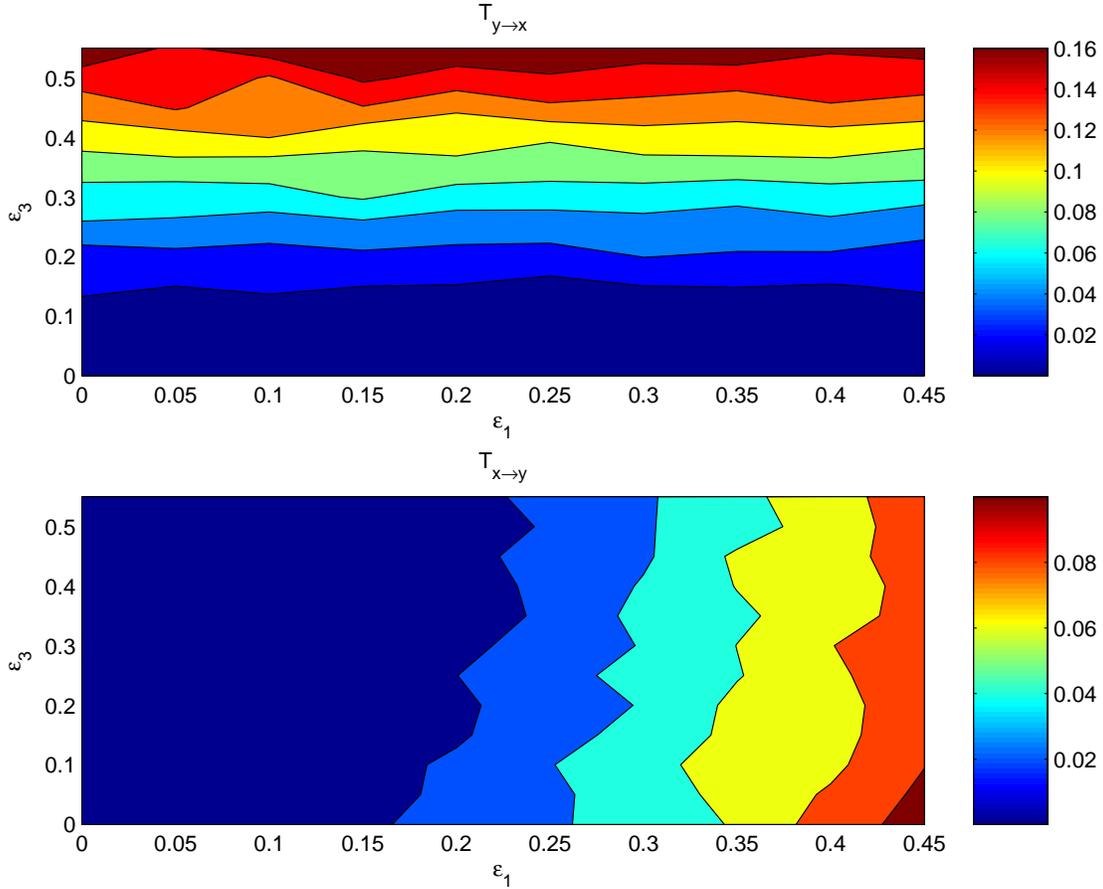}
  \caption{
	The absolute bulk information flow from subsystem $Y$ to subsystem
	$X$, and that from $X$ to $Y$. The abscissa and ordinate are
	the coupling coefficients $\varepsilon_1$ and $\varepsilon_3$,
	respectively.
	}\label{fig:T_subspace}
  \end{center}
\end{figure}

If both the coupling parameters, $\varepsilon_1$ and
$\varepsilon_3$, are turned on, the resulting causal relation 
has a distribution on the $\varepsilon_1 - \varepsilon_3$ plane.
Fig.~\ref{fig:T_individual} shows the componentwise information flows
$T_{x_1\to y_1}$ (bottom) and $T_{y_3\to x_3}$ (top) on the plane.
The other two flows, i.e., their counterparts $T_{y_1\to x_1}$ and
$T_{x_3\to y_3}$, are by computation essentially zero.
As argued in the preceding subsection, the bulk information flows should
follow the general pattern, albeit perhaps in a more coarse and/or mild
pattern, since it is a property on the whole. 
This is indeed true. Shown in Fig.~\ref{fig:T_subspace} are
the bulk information flows between $X$ and $Y$ 
computed using Eqs.~(\ref{eq:TAB_est}) and (\ref{eq:TBA_est}). 

\begin{figure}[h]
\begin{center}
  \includegraphics[width=0.9\textwidth]{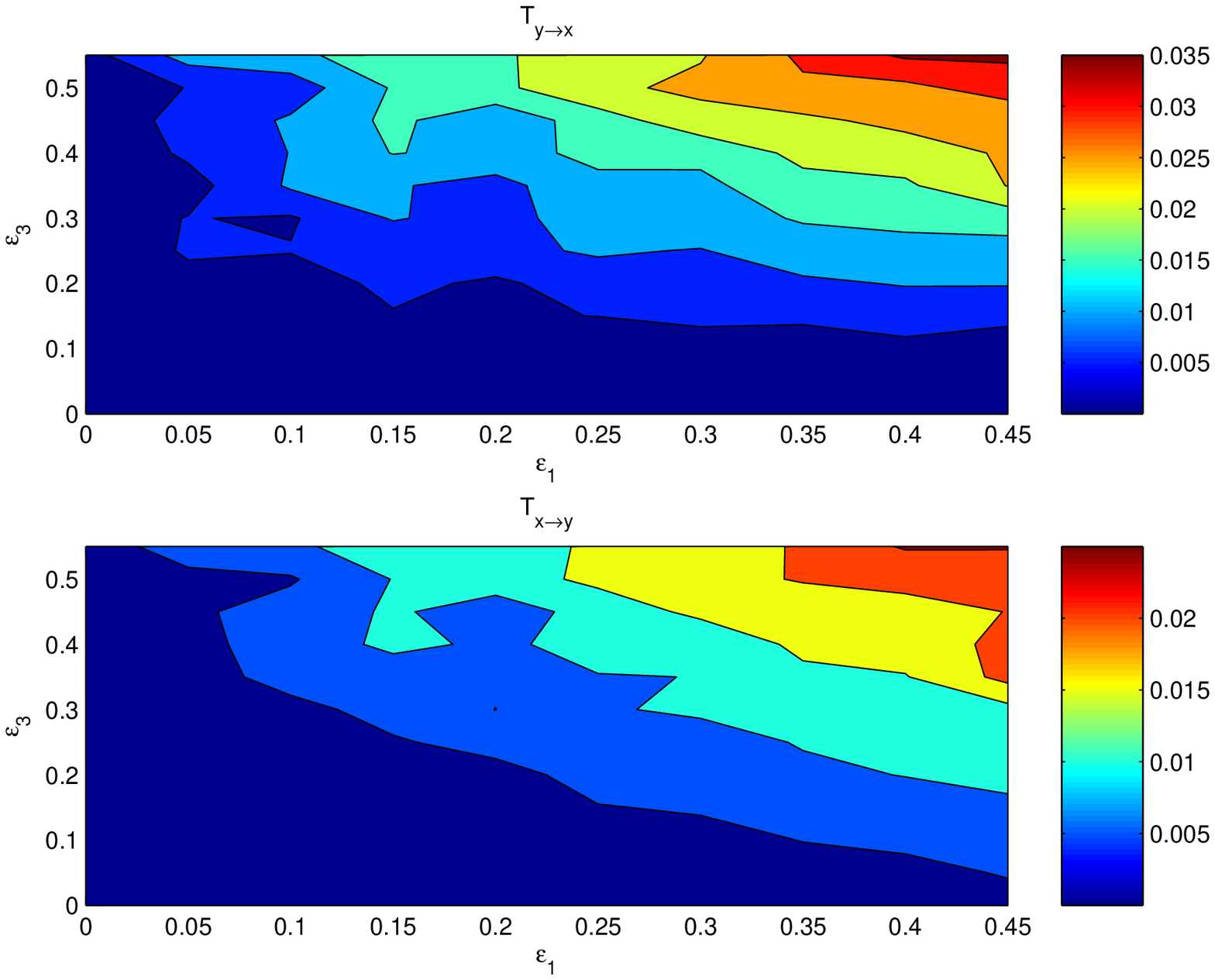}
  \caption{
	As Fig.~\ref{fig:T_subspace}, but for the information flows
	between the mean series $\bar x = \frac13(x_1+x_2+x_3)$ 
	and $\bar y = \frac13(y_1+y_2+y_3)$.
	}\label{fig:T_avg}
  \end{center}
\end{figure}

Again, as usual we try the averages and first PCs as proxies for estimating
the causal interaction between $X$ and $Y$. Fig.~\ref{fig:T_avg} shows the
distributions of the information flows between $\bar x$ and $\bar y$.
The resulting patterns are totally different from what 
Fig.~\ref{fig:T_individual} diplays; obviously these patterns are incorrect.

\begin{figure}[h]
\begin{center}
  \includegraphics[width=0.9\textwidth]{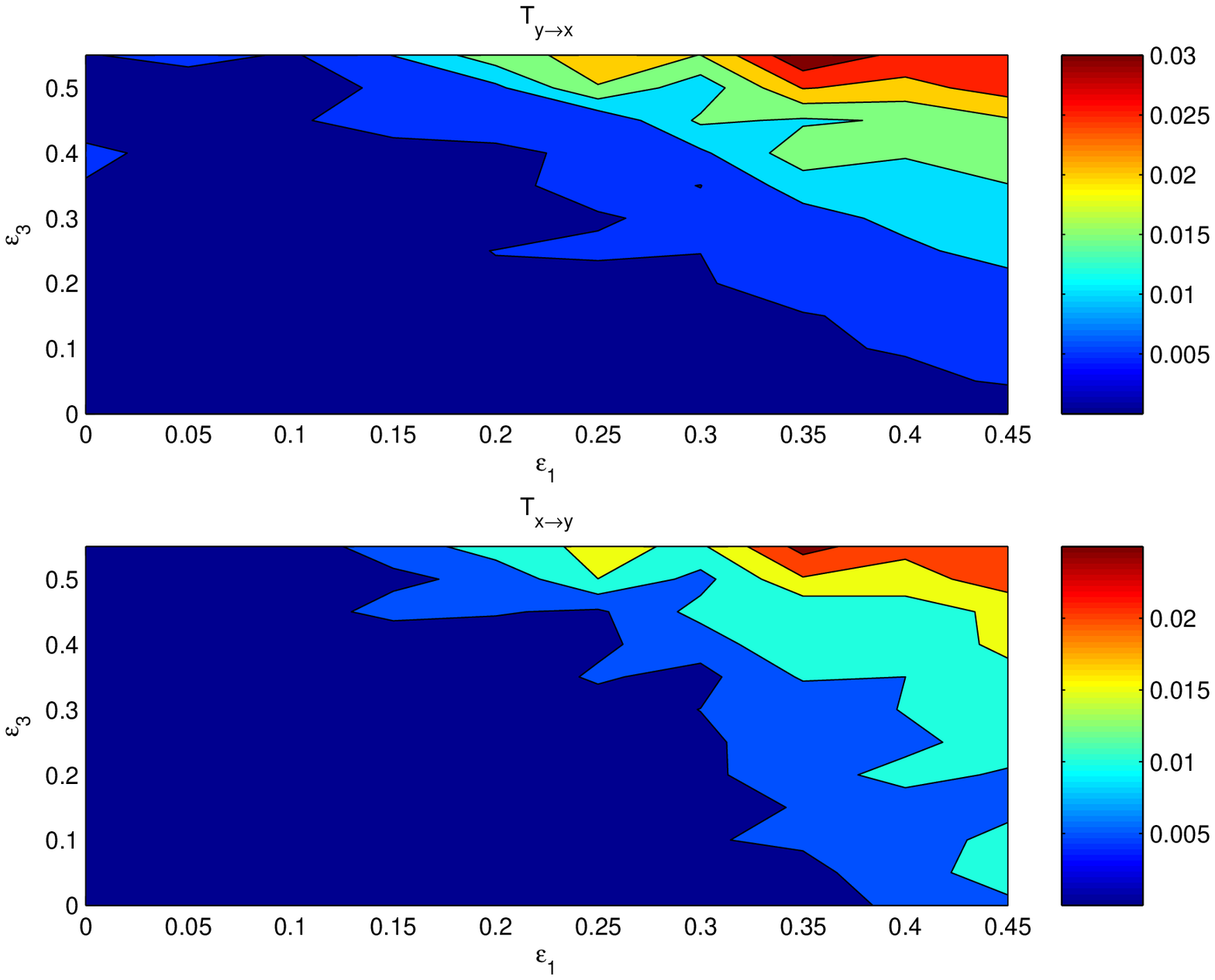}
  \caption{
	As Fig.~\ref{fig:T_subspace}, but for the information flows
	between the first principal component of $(x_1,x_2,x_3)$ 
	and that of $(y_1,y_2,y_3)$.
	}\label{fig:T_pc1}
  \end{center}
\end{figure}

One may expect that the PCA method should yield more reasonable causal 
patterns. We
have computed the first PCs for $(x_1,x_2,x_3)$ and $(y_1,y_2,y_3)$,
respectively, and estimated the information flows using the algorithm by
Liang\cite{Liang2021a}. The resulting distributions, however, are no better than
those with the averaged series. That is to say, this seemingly more 
sophisticated approach does not yield the right interaction between 
the complex subsystems, either.

\section{Summary}	\label{sect:summary}

Information flow makes a natural measure of the causal interaction between
dynamical events. In this study, the information flows between 
two complex subsystems of a large dimensional system are studied,
and analytical formulas have been obtained in a closed form. For easy
reference, summarized hereafter are the major results.

	For an $\dms$-dimensional system 
	\begin{eqnarray*}
	\dt {\ve x} = \vve F(\ve x, t) + \vve B(\ve x,t) \dot {\ve w},
	\end{eqnarray*}
	if the probability 
	density function (pdf) of $\ve x$ is compactly supported, then
the information flows from subsystem $A$, which are made of $\ve x_\subA$, 
	to subsystem $B$ made of $\ve x_\subB$ ($1\le r<s\le\dms$), and
	that from $B$ to $A$ are, respectively (in nats per unit time), 
	\begin{eqnarray*}	
	&& T_{A\to B} 
	= -E\bracket{\sum_{i=r+1}^s \frac1 {\rho_\subB}
		\int_{\R^{\dms-s}} \Di{F_i\rho_\subBother} d\ve x_\others } \cr
	&&\qquad\ \ \ \
	  + \frac12 E\bracket{
		\sum_{i=r+1}^s \sum_{j=r+1}^s \frac 1 {\rho_\subB}
	        \int_{\R^{\dms-s}} \DiDj {g_{ij} \rho_\subBother} d\ve x_\others
			     },
	\end{eqnarray*}
	\begin{eqnarray*}
	&&T_{B\to A} 
	= -E\bracket{\sum_{i=1}^r \frac1 {\rho_\subA}
		\int_{\R^{\dms-s}} \Di{F_i\rho_\subAother} d\ve x_\others } \cr
	&&\qquad\ \ \ \
	  + \frac12 E\bracket{
		\sum_{i=1}^r \sum_{j=1}^r \frac 1 {\rho_\subA}
		\int_{\R^{\dms-s}} \DiDj {g_{ij} \rho_\subAother} d\ve x_\others
			     },
	\end{eqnarray*}
	where $g_{ij} = \sum_{k=1}^m b_{ik} b_{jk}$,
	and $E$ signifies mathematical expectation.
Given $\dms$ stationary time series, $T_{A\to B}$ and $T_{B\to A}$ can be estimated.
The maximum likelihood estimators under a Gaussian assumption are referred
to Eqs.~(\ref{eq:TAB_est}) and (\ref{eq:TBA_est}).

We have constructed a VAR process to validate the formalism. The system has
a dimension of 6, with 2 subsystems respectively denoted by $X$ and $Y$,
each with a dimension of 3. $X$ drives $Y$ via the coupling at one
component, and $Y$ feedbacks to $X$ via another. The detailed, 
componentwise causal relation can be easily found using our previous
algorithms such as that in \cite{Liang2021a}. It is expected
that the bulk information flow should in general also follow a similar 
trend, though the structure could be in a more coarse and mild fashion, as
now displayed is an overall property. The above formalism does yield such a
result. On the contrary, the commonly used proxies for subsystems, such as
averages and principal components (PCs), generally do not work. Particularly, 
the averaged series yield the wrong results in the two cases considered in
this study; the PC series do not work, either, for the mutually causal
case, though they result in a satisfactory characterization 
for the case with a one-way causality.

The result of this study is applicable in many real world problems. 
As explained in the Introduction, it will be of particular use in 
related fields of
climate science, neuroscience, financial economics, fluid mechanics, 
among others. 
For example, it helps clarify the role of greenhouse gas emission 
in bridging the climate system
and the socio-economic system (see a review in \cite{Tachiiri2021}).
Likewise, the interaction between the earth system and public health
\cite{Balbus2016} can also be studied. 
In short, it is expected to play a role in the frontier field of
complexity, namely, multiplex networks, or networks of networks 
(see the references in \cite{Agostino2014}, \cite{Kenett2015},
\cite{DeFord2019}).
We are therefore working on these applications.


\begin{acknowledgments} 
This study was supported 
by 
the National Science Foundation of China (Grant \# 41975064),
and the 2015 Jiangsu Program for Innovation Research and
Entrepreneurship Groups.
\end{acknowledgments}

\end{document}